\documentclass{article}

\usepackage{arxiv}

\usepackage[utf8]{inputenc} 
\usepackage[T1]{fontenc}    
\usepackage{hyperref}       
\usepackage{url}            
\usepackage{booktabs}       
\usepackage{amsfonts}       
\usepackage{nicefrac}       
\usepackage{microtype}      
\usepackage{lipsum}
\usepackage{graphicx}
\graphicspath{ {./images/} }
\usepackage{tabularx}

\usepackage{caption}
\usepackage{tabularx}

\title{NoSQL Databases: Yearning for Disambiguation}

\author{
  Chaimae Asaad\\
  Alqualsadi, Rabat IT Center, ENSIAS, University Mohammed V in Rabat\\ and TicLab, International University of Rabat,
  Morocco\\
  \texttt{chaimae.asaad@um5.ac.ma} \\
   \And
 Karim Ba\"ina \\
    Alqualsadi, Rabat IT Center, ENSIAS, University Mohammed V in Rabat,\\
   Morocco\\
  \texttt{karim.baina@um5.ac.ma} \\
   \And
   Mounir Ghogho\\
   TicLab, International University of Rabat \\
   Morocco \\
    \texttt{mounir.ghogho@uir.ac.ma}
}

\begin{document}
\maketitle
\begin{abstract}
The demanding requirements of the new Big Data intensive era raised the need for flexible storage systems capable of handling huge volumes of unstructured data and of tackling the challenges that traditional databases were facing. NoSQL Databases, in their heterogeneity, are a powerful and diverse set of databases tailored to specific industrial and business needs. However, the lack of theoretical background creates a lack of consensus even among experts about many NoSQL concepts, leading to ambiguity and confusion. In this paper, we present a survey of NoSQL databases and their classification by data model type. We also conduct a benchmark in order to compare different NoSQL databases and distinguish their characteristics. Additionally, we present the major areas of ambiguity and confusion around NoSQL databases and their related concepts, and attempt to disambiguate them.
\end{abstract}

\keywords{NoSQL Databases\and NoSQL data models\and NoSQL characteristics \and NoSQL Classification}

\newpage
\section{Introduction}
The proliferation of data sources ranging from social media and Internet of Things (IoT) to industrially generated data (e.g. transactions) has led to a growing demand for data intensive cloud based applications and has created new challenges for big-data-era databases. 

\cite{ontology} cites among the characteristics of new and emerging data sets: Variety (diversity of types), Velocity (rapid synthesis and flow), Volume (large size) and Veracity (i.e., biases, noise and abnormality in data). These types of datasets can range from a schema-based to a schema-less nature, and from structured to unstructured types.

Following the pressure that these characteristics created for industries, data store providers felt the need to scale to OLTP/OLAP style application loads where millions of users read and update and where read/write operations are distributed over many servers. Relational database management systems have little capability to horizontally scale to these levels, which paved the way to seek alternative solutions for such scenarios \cite{surv}.

In 1998, Carlo Strozzi coined the term NoSQL to refer to his relational database management system 'Strozzi NoSQL' which did not expose an SQL interface in contrast to its traditional counterparts. Eventually, the term was revived to mean Not Only SQL, and is currently used as an umbrella term to distinguish a newly emerged heterogeneous schema-less class of databases fit for Big Data needs and differing from relational databases in terms of technical characteristics and data model. When technology giants like Google and Amazon created their own data stores as solutions to the large amount of data they deal with and to fulfill their functional requirements, other vendors and open-source developers were inspired to do the same, thus contributing to the widespread of NoSQL solutions \cite{surv}.

Interest in NoSQL technologies continued to bloom, leading to the publication of several works, particularly by Stonebraker (\cite{stone1}, \cite{stone2}) and Cattell \cite{cattell}. By 2011, the NoSQL ecosystem was thriving, with several databases at the center of multiple studies (\cite{study1}, \cite{study2}). Research was then directed towards identifying and classifying NoSQL databases according to a set of architectural features and goals, and towards building a theoretical background for NoSQL in the hope of standardizing it.

In the midst of all the literary and technical resources targeting NoSQL databases, a lot of misconceptions became common practices. The ambiguity and nuances that characterize a newly founded academic discipline created major breaks in communication and understanding. The lack of a stable standard and a solid theoretical background, along with the ever-changing nature of computer science sub-fields left room for false interpretations around major concepts related to NoSQL databases.

In this paper, we provide a survey of NoSQL databases, their classification, their characteristics, uses and their technical advantages. We also present the data model each NoSQL family is designed after, and clarify ambiguities surrounding them. Additionally, we present major areas of ambiguity around NoSQL databases and their related concepts and attempt to clarify them.

The remainder of this paper is organized as follows. Section 2 

\section{Characteristics and uses of NoSQL Databases}
\label{sec:headings}
NoSQL databases are used nowadays by cloud application developers to harness large amounts of shema-less data. They have a flexible schema, implying the possibility of a dynamically evolving schema. NoSQL databases do not require schema definition prior to inserting data nor do they demand schema modification when data management needs evolve \cite{surv}.

The primary uses of NoSQL Databases include large-scale parallel data processing over distributed systems, embedded machine-to-machine information look-up and retrieval, exploratory analytics on semi-structured data and large volume data storage ranging from small-packet structured to semi-structured and unstructured data. NoSQL databases provide relatively inexpensive, highly scalable storage for high-volume, small-packet historical data like logs, call-data records, meter readings and ticker snapshots, and for unwieldy semi-structured or unstructured data (e.g. email archives, xml files, documents). Their distributed framework also makes them ideal for massive batch data processing such as aggregating, filtering, sorting and other algorithmic crunching scenarios. They are also a good fit for machine-to-machine data retrieval and exchange, and for processing high-volume transactions. These systems are also very good exploratory analytics against semi-structured or hybrid data \cite{Moniruzzaman}.
There are various reasons behind the search for alternate solutions to Relational databases. The rich feature set and the ACID properties implemented by RDBMSs might be necessary for specific use cases. However, the cost associated with the scaling of increasing volumes of relational database management systems is very expensive. In contrast, NoSQL data stores are designed to scale well horizontally and run on commodity hardware. In addition, the 'one size fits all' notion does not fit the current application scenarios and it is better to build systems based on the nature of the application and its work/data load. Moreover, the needs for data storage have increased over the years. RDBMSs were designed for large high-end machines and centralized deployments, however, today's companies use commodity hardware in a distributed environment. Today's data is not rigidly structured and does not require dynamic queries \cite{surv}.

NoSQL databases present many advantages over traditional ones, among which \cite{ontology} mentions:High concurrent reading and writing of data with low latency in cloud specific systems; String scalability and availability;Lower operational and management cost; The ability to replicate and scatter data over over many servers; Efficient use of distributed indexes and RAM for data storage; The ability to dynamically add new attributes in data records; Flexible schema support; and Compatibility with the CAP (Consistency, Availability and Partition tolerance) model and BASE (Basic Availability, Soft state, Eventual Consistency) principles to comply with distributed systems characteristics.

In order to guarantee the integrity of data, most of the classical database systems are based on transactions. This ensures consistency of data in all situations of data management. These transactional characteristics are also known as ACID (Atomicity, Consistency, Isolation and Durability). However, scaling out (i.e., scaling horizontally or adding more nodes to a system) of ACID-compliant systems has proven to be a problem. Conflicts are arising between the different aspects of high availability in distributed systems that are not fully solvable – known as the CAP-theorem \cite{Moniruzzaman}.

In a keynote titled 'Towards Robust Distributed Systems" \cite{brewer} at ACM’s symposium on Principles of Distributed Computing in 2000, Eric Brewer introduced the CAP-theorem which is widely adopted today by large web companies (e.g. Amazon) as well as in the NoSQL research community.

The CAP acronym stands for (as cited in \cite{strauch}):\\
\begin{itemize}
\item[-] \textbf{Consistency:} meaning if and how a system is in a consistent state after the execution of an operation. A distributed system is typically considered to be consistent if after an update operation of some writer all readers see his updates in some shared data source. There are nevertheless several alternatives towards this strict notion of consistency.\\
\item[-] \textbf{Availability:} and especially high availability, meaning that a system is designed and implemented in a way that allows it to continue operation (i.e., allowing read and write operations) if for instance nodes in a cluster crash or some hardware or software parts are down due to upgrades.\\
\item[-] \textbf{Partition tolerance}: understood as the ability of the system to continue operation in the presence of network partitions. These occur if two or more "islands" of network nodes arise, temporarily or permanently, which cannot connect to each other. Some also understand partition tolerance as the ability of a system to cope with the dynamic addition and removal of nodes (e.g. maintenance purposes).\\
\end{itemize}

The CAP theorem postulates that \cite{strauch} in a "shared-data system", only two of the three CAP characteristics can be achieved fully at the same time. For a growing number of applications and use-cases (including web applications, especially in large and ultra-large scale, and even in the e-commerce sector), availability and partition tolerance are more important than strict consistency. These applications have to be reliable which implicates availability and redundancy (consequently distribution among two or more nodes, which is necessary as many systems run on cheap, commoditized and unreliable machines and also provides scalability).
These properties are difficult to achieve with ACID properties, therefore approaches like BASE are applied. Other properties relating to NoSQL technologies include, among others, sharding (i.e., Horizontal partitioning by some key and storing records on different servers in order to improve performance), Horizontal scalability (i.e., Distributing both data and load over many servers) and Vertical scaling (i.e., Use of multiple cores and/or CPUs by a DBMS) \cite{strauch}.\\

The BASE approach according to Brewer \cite{brewer} forfeits the ACID properties of consistency and isolation in favor of “availability, graceful degradation and performance”. The acronym BASE is composed of the following characteristics: Basically available; Soft-state; Eventually consistent.
With regards to databases, Brewer \cite{brewer} concludes that current databases are better at consistency than availability, and that wide-area databases can’t have both–a notion that is widely adopted in the NoSQL community and has influenced the design of non-relational data stores. Systems that can be characterized by the BASE properties include Amazon’s Dynamo\cite{sadalage}, which is available and partition-tolerant but not strictly consistent, i.e writes of one client are not seen immediately after being committed to all readers. Google’s BigTable \cite{chang} chooses neither ACID nor BASE but the third CAP-alternative being a consistent and available system and consequently not able to fully operate in the presence of network partitions \cite{strauch}. 

Brewer \cite{brewer} points out traits and examples of the three different choices that can be made according to the CAP-theorem. Additionally, Brewer contrasts ACID with BASE while considering the two concepts as a spectrum instead of alternatives excluding each other. \cite{ippolito} summarizes the BASE properties in the following way: an application works basically all the time (basically available); does not have to be consistent all the time (soft-state); but will be in some known-state state eventually (eventual consistency).
\section{Modern CAP: Disambiguation}
In the twenty years since its proposal, the CAP theorem has been widely adopted by the database community. However, it has also remained a subject of constant debate, misunderstandings, ambiguity and, at times, even criticism.
\subsection{The Consistency-Availability Trade-off}
The CAP theorem was formulated to describe a major characteristic of distributed systems. In an era where scalability out of ACID properties was deemed inefficient, the CAP theorem was used as an argument in favor of adopting NoSQL databases. A rising number of database specialists have seeked to debunk this argument based on the fact that it is now possible for scalability to be maintained along with a strong consistency (e.g. NewSQL systems), as is stated in \cite{distrib}.

For NoSQL Databases, the argument holds still. The scalability provided by NoSQL comes at the 'small' cost of having to prioritize availability over consistency in the existence of a network partition. Evidently, when the system is functioning in its normal state, the database can and should be able to provide both characteristics.

In 2012, Eric Brewer wrote a follow-up paper entitled "CAP Twelve Years Later: How the 'Rules' Have Changed" \cite{brewer12}. In it, Brewer explained the misuse of the CAP theorem by many designers and researchers over the years. The typical CAP definition states that "a shared-data system can't have all 3". This definition has stirred misunderstandings about what the CAP theorem is and what it means because it implies that some systems are unavailable or inconsistent 100\% of the time. Brewer explains in \cite{brewer12} that the easiest way to understand CAP is to think of two nodes on opposite sides of a partition. \\
- \textit{Scenario 1:} Allowing at least one node to update state will cause the nodes to become inconsistent, thus forfeiting Consistency.\\
- \textit{Scenario 2:} If the choice is to preserve consistency, one side of the partition must act as if it is unavailable thus forfeiting Availability.

Evidently, consistency requires communication between all the nodes of the distributed system. Only when nodes communicate (normal state) is it possible to preserve both consistency and availability, thereby forfeiting Partition tolerance (P). However, the general belief is that for wide-area systems, designers cannot forfeit (P) \cite{brewer12}. 
\begin{table}[ht]
\caption{CAP-theorem: Alternatives, traits and examples \cite{strauch}}
\begin{tabular}{  | m{5cm} | m{5cm}| m{5cm} | } 
\hline
\textbf{Choice} & \textbf{Traits} & \textbf{Examples} \\ 
\hline
Consistence  +\newline Availability \newline (Forfeit Partitions) & 2-phase-commit,\newline Cache-validation protocols & Single-site databases, Cluster databases, LDAP,\newline xFS system \\ 
\hline
Consistency + \newline Partition tolerance\newline (Forfeit Availability) & Pessimistic locking, Make minority partitions unavailable & Distributed databases, Distributed locking, Majority protocols \\ 
\hline
Availability + \newline Partition tolerance \newline (Forfeit Consistency) & Expiration/leases, Conflict resolution, optimistic & Coda,\newline Web caching, \newline DNS\\
\hline
\end{tabular}
\end{table}
CAP is basically an availability vs. consistency trade-off. In practice, a distributed system cannot avoid partitions. A distributed database which satisfies the (CA) CAP characteristics is not a valid theoretical or practical possibility. However, it is believed that, for a distributed application, a (CA) database can be built in the form of a relational database (e.g. PostgreSQL) deployed to multiple nodes using replication. This is a work-around to provide a distributed architecture for (CA) relational databases using various methods (e.g. Microsoft Citus Data, Postgres-XL). As aforementioned, a new type of databases (NewSQL databases) has emerged seeking to provide that same characteristic (NoSQL-level scalability) while remaining ACID-compliant.
ACID and BASE represent two design philosophies at opposite ends of the Consistency-Availability spectrum.
In some ways, the NoSQL movement is about creating choices that focus on availability first and consistency second. Soft state and eventual consistency are techniques that work well in the presence of partitions and thus promote availability. BASE was created to capture the emerging '90s design approaches for high availability and to make explicit both the choice and the spectrum. Databases that adhere to ACID properties focus on consistency and represent the traditional approach. Modern large scale wide-area systems (including the cloud) use a mix of both approaches \cite{brewer12}.

The relationship between ACID and CAP is somewhat complex and often misunderstood in part because C and A in ACID and CAP are not equivalent. 

\begin{table}[h]
\caption{CAP and ACID}
\begin{tabular}{  | c | m{7cm}| m{7cm} | }
\hline
\textbf{Characteristic} & \textbf{CAP} & \textbf{ACID} \\
\hline
\textbf{A} & \textbf{Availability:} \newline All live nodes in a distributed system can process operations and respond to queries \cite{volt1}, i.e., any replica has to reply to any received request \cite{distrib}.

& \textbf{Atomicity:} \newline All components of a transaction are treated as a single action. All are completed or none are; if one part of a transaction fails, the database’s state is unchanged \cite{volt1}.

\\
\hline
\textbf{C} & \textbf{Consistency:} \newline All nodes see the same data values at the same time, i.e., each read request returns the last written value. This property corresponds to linearizability (consistency over individual operations) and not serializability (consistency over groups of operations) \cite{distrib}.
& \textbf{Consistency:} \newline Transactions must follow the defined rules and restrictions of the database, e.g., constraints, cascades, and triggers. Thus, any data written to the database must be valid and any transaction that completes will change the state of the database. No transaction can create an invalid data state \cite{volt1}.\\
\hline
\end{tabular}
\end{table}

Brewer \cite{brewer12} further explains the complicated nuances in the relationship between ACID properties and the CAP theorem. He explains that choosing availability affects only some of the ACID guarantees. 
In the case of Atomicity and Durability, there is no absolute need to forfeit them in a CAP system. When the focus is availability, both sides of the partition should still use atomic operations, and even though the developer may opt out of a durability choice given its expense, there is no mandatory forfeiting rule. Brewer also states that isolation is at the core of CAP in the sense that if the system requires ACID isolation, it can operate on at most one side during a partition given that serializability requires communication in general and thus fails across partitions. 

In general, running ACID transactions on each side of a partition makes recovery easier \cite{brewer12}. It is important to also note that ACID properties were gathered to create an acronym and that they are not an all-or-nothing rule, but rather a set of rules that can apply individually and in groups.

\subsection{Partition Tolerance}
"Scalability is a desirable attribute of a network, system or process. The concept connotes the ability of a system to accommodate an increasing number of elements or objects to process growing volumes of work gracefully, and/or be susceptible to enlargement" \cite{bondi}. Historically, database scalability started with the aim of providing service on mainframes, minicomputers and personal computers. Once software limitations and critical scalability bottlenecks were resolved, hardware-oriented innovations took place (multiprocessor computers, multi-core processors, etc.). 

Basically, scalability can be achieved following one of the two following paradigms. Vertical scalability (scaling up) is adding resources to a single node in a system, typically involving the addition of processors or memory to a single computer. Horizontal scalability (scaling out) involves adding more nodes to a system, such as adding a new computer/server to a distributed system \cite{abbadi}. Horizontal scaling is supposedly more efficient and dynamic whereas vertical scaling is generally limited to the capacity of a single machine which adds upper limit and downtime constraints. NoSQL systems gradually gained favor over traditional databases thanks to their horizontal scalability capacities and their flexible data models. This scalability was accompanied by a loosening of consistency requirements in favor of eventual consistency.

Two philosophies on opposite sides of the database clustering spectrum were proposed: (i) Allowing distributed transactions to affect data stored on separate computes thus establishing the shared-nothing architecture \cite{solidfire}, and (ii) providing full functionality on multi-server clusters using the shared-everything architecture introduced Oracle. 

Similarly to most distributed systems, the shared-nothing architecture is a common design choice for NoSQL databases given its particular effectiveness for them. The scalability of shared-nothing clustering makes it ideal for read-intensive analytical processing typical of data warehouses. Shared-nothing systems are comparatively
cheaper to build because they do not require any special hardware and can use commodity hardware all while achieving high reliability through redundancy across multiple geographically distributed datacenters \cite{distrib}.

In a shared-nothing architecture, each node is made of processor, main memory and disk and communicates with other nodes through the interconnection network. Each node is under the control of its own copy of the operating system and thus can be viewed as a local site (with its own database and software) in a distributed database system. Therefore, most solutions designed for distributed databases such as database fragmentation, distributed transaction management and distributed query processing may be reused \cite{ency}. 

These systems, given their distributed nature, are prone to partitions. Although  reliable networks make them a rare event, they are still a consideration to be planned for, which is why many database specialists criticized the CAP theorem for implying that Partition tolerance can be sacrificed. 

A main property often considered a disadvantage of shared-nothing architecture is that a partitioning scheme must be designed. A source of ambiguity (especially for novices and students) around the (P) in the CAP theorem revolves around differentiating between the completely unrelated concepts of network \textbf{\textit{partitions}} and data \textbf{\textit{partitioning}}. 

In the context of distributed systems, a \textbf{\textit{partition}} refers to a network split between nodes causing communication errors within the system. The process of dealing with partitions includes detecting them through a quorum on available instances \cite{stone10}. After a network failure, there are basically two ways to propagate an update on a node. If the propagation is done asynchronously, the update will have to wait and an "island" or a node will have to stay inconsistent. If the propagation is done synchronously, the entire system will not be available until the update is propagated \cite{nicome}. Consequently, and as aforementioned, in both of these situations, either consistency or availability will have to be sacrificed.
Simply put, Partition tolerance implies that the system must be able to continue functioning in the presence of partitions. A partition is a break in communication within a distributed system. 
 
On the other hand, \textbf{\textit{partitioning}} is a way of intentionally breaking a large database down into smaller ones. The main reason for wanting to partition data is scalability. Different partitions can be placed on different nodes in a shared-nothing cluster. Automatic partitioning is known as sharding. A partition is known as: a shard (MongoDB, Elasticsearch, and SolrCloud); a region (HBase), a tablet (Bigtable), a vnode (Cassandra and Riak), and a vBucket (Couchbase). Partitioning is usually combined with replication to ensure fault tolerance \cite{distrib}. 

Some ambiguity also surrounds differentiating between timeouts, dead nodes, slow nodes and partitioned nodes (in the context of network partitions). In a recent podcast (\cite{brewerpodcast}) , Eric Brewer explained that a network partition cannot be reliably detected, and that even if the timeout expires for whatever reason, leaving operations unable to be completed, only two choices are available to achieve consistency: delaying, which leads to unavailability; or "giving up" thus forfeiting consistency. 

As for the difference between a partitioned node, a dead node and a slow node, it is important to state that in its formalized  (Gilbert and Lynch) text, availability requires only non-failing nodes to respond. Failed machines are inherently not expected to respond to clients' requests. For an asynchronous network, messages can be arbitrarily delayed, which entails that it is difficult (even impossible) to distinguish between failed messages and delayed ones in a finite amount of time. 
This particular ambiguity brings up discussions around the differences between FLP theorem (also known under the consensus impossibility result and the Fischer, Lynch, and Paterson theorem) and the CAP theorem. FLP or consensus is a fundamental problem in distributed computing. The impossibility result states that, "in a fully asynchronous message-passing distributed system, in which one process may have a crash failure, consensus is impossible" \cite{flp}. The main difference between the theorems, is that FLP deals with delayed but not lost messages, whereas CAP states that messages may be lost. Another distinction arizes around the consensus problem, specifically how what is considered a partitioned node in CAP is actually a failed node in FLP and consequently is exempt from having to achieve consensus (\cite{trail1}, \cite{trail2}).

\subsection{CAP Criticism and Alternatives}

The previous sections seeked to disambiguate common misinterpretations of the CAP theorem in terms of the "2 out of 3" rule and it failed to capture the true nuances of the trade-off reality in distributed databases. As an impossibility result, the CAP theorem has received criticism over the years with many researchers proposing improved versions and alternatives. In what follows, we will present those proposed alternatives and the facets of CAP which face the most criticism.

Consistency, in a distributed database context, is defined by how copies of the same data vary within the same replicated database system. Consistency can be seen from two aspects: data-oriented consistency and client-oriented consistency. \cite{consist}. As aforementioned, following the BASE properties for NoSQL databases implied prioritizing (high) availability during a network partition, while being able to provide various forms of consistency. For a long time, CAP has been criticized for formulating a binary view on availability and consistency when the reality is that those properties can vary in levels. 12 years after the CAP theorem was first proposed, Brewer explained in \cite{brewer12} that "all three properties are more continuous than binary. Availability is obviously continuous from 0 to 100 percent, but there are also many levels of consistency, and even partitions have nuances, including disagreement within the system about whether a partition exists".

NoSQL storage systems provide various consistency models, depending on how they were implemented and the priorities they were set to handle. Types of consistency models include: Strong consistency, weak consistency, eventual consistency, causal consistency, read-your-writes consistency, session consistency, monotonic reads consistency, and monotonic writes consistency \cite{consist}. Other systems (e.g. Yahoo Sherpa, Cassandra) offer tunable configurations for consistency in order for users to choose which consistency model (and consequently availability and latency level) is best suited for their application.

Another major criticism of CAP discusses how the the theorem does not mention a very important system property: Latency.  \cite{abadiblog} and \cite{kleppblog} raise the issue of latency being overlooked in the CAP theorem thus reducing its utility and its application in real-life cases for distributed systems. \cite{abadiblog} proposes an improvement over the CAP theorem: PACELC, i.e., if there is a Partition, how does the system handle the Availability-Consistency trade-off; Else, when the system is in its normal state (absence of partitions), how does the system tradeoff between Latency and Consistency. 
A formal proof of PACELC was presented recently in \cite{proof}.

The CAP theorem has also been heavily criticized for only considering one kind of fault (i.e., network partitions) \cite{kleppman} and for missing the nuances behind real error scenarios where the theorem does not apply \cite{stone10}.

In \cite{kleppman} and \cite{critickleppman}, Pr. Martin Kleppman states that the CAP theorem only considers the case where nodes are alive but disconnected from each other, and that "using network partition separately in CAP as a particular type of fault only creates confusion". \cite{kleppman} also adds that the CAP theorem does not offer any trade-offs for network delays, dead nodes, etc. Kleppman states that "although CAP has been historically influential, it has little practical value for designing systems". \cite{kleppman} considers that "there are more interesting impossibility results in distributed systems \cite{lynch}, and that CAP has been superseded by more precise results (\cite{attiya}, \cite{ajoux})".

\section{Classification of NoSQL Databases}

NoSQL has been used as an umbrella term to refer to various relatively new databases characterized by their schema-less nature, their capacity to handle Big Data and their rejection of the 'one size fits all' paradigm. They also use (often but not always) query languages other than SQL. As aforementioned, the CAP theorem states that, in the presence of a network partition, a distributed database can achieve either consistency of availability. Following this theorem, a popular classification of NoSQL databases, known as the cap triangle, is used to distinguish CP, AP and CA systems. The first flaw of this classification is that (CA) systems are single-node databases and thus should not be of concern in the CAP context, since Partitions simply cannot be "sacrificed" or opted out of in a distributed system \cite{coda}. Another flaw in this logic is what Kleppman mentions in \cite{kleppblog}, where he states that "so far we haven’t been able to rigorously classify them as “AP” or “CP”, either because it depends on the particular operation or configuration, or because the system meets neither of the CAP theorem’s strict definitions of consistency or availability".

In order to capture the nuances and enrich this classification, we will present in what follows, a consensus-based data-model-oriented classification of NoSQL databases, along with their characteristics and where they fall on the availability-consistency spectrum in terms of which consistency model they implement and what trade-offs do they apply under different circumstances and following various application needs. This perspective permits to view the classification of NoSQL databases from a less binary lens, and aims to enrich this survey.

\subsection{Data-model Based Classification of NoSQL Databases}

A 'first' classification of NoSQL databases came from Leavitt \cite{Leavitt} who classified NoSQL databases in three 'popular' types: Key-Value Stores (e.g. Amazon’s SimpleDB, Uppsala University’s Amos II, Scalaris); Column-oriented databases (e.g. Facebook's Cassandra, Apache Software Foundation's Hbase); Document-based stores (e.g. Apache Software Foundation's CouchDB, MongoDB, Basho Technologies’ Riak). This classification, although accepted in the database community, is not the most recent or agreed upon. Instead, the consensus-based classification involves the three aforementioned types and adds Graph-oriented databases. In what follows, we will specify the characteristics of each type and cite the databases it includes. We will also disambiguate various concepts surrounding these types. 

\subsubsection{Key-value Databases}

Key-Value databases are implemented using a Hash Table containing an alpha-numeric identifier serving as a unique key and a pointer specifying the particular item of data which value is associated with the key. The result of this mapping creates a key-value pair. The Value may be simple text strings or more complex lists and sets. Hash Tables are useful for searches of simple or complex values in extremely large datasets \cite{kv1}. 

Key-Value stores are similar to maps or dictionaries where data is addressed by a unique key. Since values are uninterpreted byte arrays, which are completely opaque to the system, keys are the only way to retrieve stored data. Values are isolated and independent from each other, consequently, relationships must be handled in application logic. Due to this very simple but efficiently flexible data structure, key value stores are completely schema-free. A new value of any kind can be added at runtime without conflicting any other stored data and without influencing system availability \cite{kv2}. 

Grouping key value pairs into a collection is an alternative possibility which can provide, to some extent, structure to the data model. 

Key-Value databases can handle a very large number of records. They can support high volumes of state changes per second with millions of simultaneous users through distributed processing and storage. They rely on their redundancy to face the loss of storage nodes and to protect applications. Key-Value stores are useful for simple operations, which are based on key attributes only. For instance, an order to speed up a user specific rendered webpage, parts of this page can be calculated before and served quickly and easily out of the store by user IDs when needed \cite{kv1}.

Since most Key-Value stores hold their dataset in memory, they are often used for caching of more time intensive SQL queries.  They are very useful for both storing the results of analytical algorithms and for producing those results via reports \cite{kv2}. 

Along with the number of advantages Key-Value Databases demonstrate, they have inherited a number of drawbacks from their NoSQL nature. Firstly, they do not provide any kind of traditional database capabilities, which forces them to rely on the application itself to ensure transactions. Secondly, users cannot access data by value. It is indeed impossible to query a key-value datastore in order to extract all records containing a particular set of value, and the only way to do so would be by specifying a request by key or by range of keys \cite{kv3}.

The simplicity of Key-Value stores makes them ideally suited for lightning-fast, highly-scalable retrieval of the values needed for application tasks like managing user profiles or session information or retrieving product names. They are also very useful in e-commerce, for instance, to manage shopping cart contents, product categories and reviews, etc. In networking and data maintenance, key-value datastores can be used to store Internet Protocol (IP) forwarding tables, data directories and even entire web pages by using the URL as the key and the web page as the value \cite{ian}.

Examples of Key-Value databases include \textbf{Redis}, \textbf{Riak} and \textbf{Voldemort}.

\textbf{Redis} is an in-memory, key-value datastore. It is widely used by IT brands and businesses around the world. Amazon Elastic Cache supports Redis, which makes it a very powerful and must-know key-value database. Primary memory is limited in size and more expensive than secondary ones, therefore Redis cannot store large files or binary data. However, it can store less substantially sized textual information which needs to be accessed, modified and inserted at a very fast rate \cite{redis}.
Redis Client and Server can be in the same computer or in two different ones. Redis Server is responsible for storing data in-memory. It handles all kinds of management tasks and forms the major part of the architecture. Redis Client can be Redis' console client or any other proramming language's Redis API (Application Programming Interface). Given Redis' in-memory storage, and given that the primary memory is volatile, datastore persistence is absolutely necessary for Redis, otherwise all stored data would be lost once Redis Server or Client are restarted. Basically, there are three techniques to ensure Redis' persistence. The \textit{RDB} (Redis Database file) mechanism makes a copy of all the data in-memory and stores it in secondary storage, which is permanent. This copying only happens in specific intervals, which entails that some data might be lost if they were set after RDB's last snapshot. \textit{AOF} (Append Only File) logs all the write operations received by the server, which ensure persistence. However, AOF's mechanism of writing in disk every operation makes it expensive and results in a much larger file than RDB. \textit{SAVE command} forces Redis server to create an RDB snapshot anytime using the Redis console client's Save command. AOF and RDB can be used simultaneously to get the best persistence possible. Redis does not provide any mechanism for datastore backup and recovery. Therefore, if there is any hard disk crash, all data will be lost. However, if Redis is used in a replication environment, there is no need for backup. In a replication environment, many computers share the same data with each other so that even if a few go down, all the data will still be available. This technique enables fault-tolerance and data accessibility \cite{redis}.
The Master and slaves are all Redis servers. The slaves contain exactly the same data as the master server. When a new slave is inserted in the environment, the master automatically syncs all data to the slave. Replication helps avoid disk failures and other kinds of hardware failures. In addition, it helps execute multiple read/sort queries simultaneously. It is worth noting that using persistence and replication tgether ensures that all data will be safe and protected from unexpected failures (e.g. whole replicated environment goes down due to power failure). Clustering is a technique by which data can be sharded into many nodes, with the main advantage of storing more data (because a cluster is a combination of computers). Each node is a Redis server configured as a cluster node. Persistence mechanisms should however be used, otherwise, if one node fails, the whole cluster stops working \cite{redis}.

\textbf{Riak} KV is a highly resilient NoSQL Database. It ensure that the most critical data is always available and that Big Data applications can scale. Riak can be operationalized at lower costs than both relational and other NoSQL databases, especially at scale. Running on commodity hardware, Riak is operationally easy to use with the ability to add and remove capacity on demand without data sharding or manually restructuring clusters. At its core, Riak is a Key-Value database built to distribute data across a cluster of servers, called nodes. A Riak cluster is a group of nodes that are in constant communication in order to ensure data availability and partition tolerance. Riak has a masterless architecture in which every node in a cluster is capable of serving write/read requests. All nodes are homogeneous with no single master or point of failure. Any node selected can serve an incoming request, regardless of data locality, providing data availability even when hardware, or the network itself, is experiencing failure conditions \cite{riak1}. Riak automatically distributes data across nodes in a Riak cluster and yields a near-linear performance increase as capacity is added. Data is distributed evenly across nodes using consistent hashing. When new machines are added, data is rebalanced automatically in a non-blocking operation. Since replication improves availability and partitions allow an increase in capacity, Riak combines both partitions and replications to work together. Data is partitioned as well as replicated across multiple nodes to create a horizontally scalable system. By default, Riak KV replicas are eventually consistent (i.e., even though data is always available, replicas may have the most recent update at the exact same time, causing brief periods of inconsistency while all state changes are synchronized) \cite{riak1}.

\textbf{Voldemort} is a distributed key-value storage system, where data is automatically replicated over multiple servers and partitioned so each server contains only a subset of the total data. Voldemort provides tunable consistency, and a transparent handling of server failure. It offers a number of pluggable storage engines, in addition to pluggable serialization. Data items are versioned to maximize data integrity in failure scenarios without compromising the availability of the system. Each node is independent of the other nodes, thus eliminating any central point of failure of coordination. Depending on the machines, the network and the disk system and data replication factor, the average number of operations is expected to be around 10-20k/s. Voldemort is used at LinkedIn by numerous critical services powering a large portion of the site. Contrary to relational databases, Voldemort does not attempt to satisfy arbitrary relations while satisfying ACID properties. Nor does it, like Object databases, attempt to transparently map object reference graphs. Nor does it introduce any abstraction such as document-orientation. Basically, Voldemort is just a big, distributed, persistent, fault-tolerant hash table \cite{hewhomustnotbenamed}.

\subsubsection{Column-Family Databases}

Column Family Stores are also known as column-oriented stores, extensible record stores and wide columnar stores. They were inspired from Google's BigTable, a distributed storage system for managing structured data that is designed to scale to a very large size. BigTable is used in many Google projects varying in requirements of high throughput and latency-sensitive data serving. In this map, an arbitrary number of key-value pairs can be stored within rows. Since value cannot be interpreted by the system, relationships between datasets and any other data type (other than string) are not supported natively. Multiple versions of a value are stored in chronological order to support versioning on the one hand and achieving better performance and consistency on the other one. Columns can be grouped to column families, which is especially important for data organization and partitioning. Columns are rows can be added very flexibly at runtime but column families have to be predefined oftentimes, which leads to less flexibility than the one offered by key value stores and document stores. Due to their tablet format, column-family databases have a similar graphical representation compared to relational databases. The main difference lies in their handling of null values. Considering a use case with many different kinds of attributes, relational databases would store a null value in each column the dataset has no value for. In contrast, column-family stores only store a key-value pair in one row, if the dataset needs it. This characteristic is what Google describes as ''sparse'' and which makes column-family stores very efficient in domains with huge amounts of data with varying numbers of attributes. Hbase and HyperTable are open source implementations of BigTable, whereas Cassandra differs from their data model, since another dimension called supercolumns is added. These supercolumns can contain multiple columns and can be stored within column families. Therefore, Cassandra is more suitable for handling complex and expressive data structures \cite{kv1}.

\textbf{HBase} is an open source NoSQL column-family database that runs on top of HDFS . It provides read/write access to large datasets, and scales linearly to handle huge data sets with billions of rows and millions of columns, and it easily combines data sources that use a wide variety of different structures and schemas. HBase is natively integrated with Hadoop and works seamlessly alongside other data access engines through YARN (Yet Another Resource Manager). Apache HBase provides random, real real time access to data in Hadoop. It was created to host very large tables, making it a great choice to store multi-structured or sparse data. Users can query HBase for a particular point in time, making 'flashback' queries possible. HBase has many characteristics, it is fault-tolerant, which gives it the advantages of replication across the data center, atomic and strongly consistent row-level operations, high availability through automatic failover, and automatic sharding and load balancing of tables. HBase is also fact and usable, it provides near real-time lookups, in-memory caching and server-side processing via filters and co-processors and its data model accomodates a wide range of use cases \cite{hbase2}.

\textbf{HyperTable} is a high performance, open source, massively scalable database modeled after Bigtable, Google’s proprietary scalable database managment system. Hypertable is designed to utilize a scalable and highly available file system such as Hadoop (HDFS), where high availability is achieved by replicating file data across multiple machines. The \textit{Master} handles all Meta operations such as creating and deleting tables. Client data does not move through the Master, so the Master can be down for short periods of time without clients being aware. The master is also responsible for detecting range server failures and reassigning ranges if necessary. \textit{Range Servers} are responsible for managing ranges of table data, handling all reading and writing of data. They can manage up to potentially thousands of ranges and are agnostic to the set of ranges that they manage or the tables of which they’re a part. Ranges can move freely from one range server to another, an operation that is mostly orchestrated by the Master. Hypertable is capable of running on top of any filesystem. To achieve this, Hypertable has abstracted the interface to the filesystem by sending all filesystem requests through a Distributed File System (DFS) broker process. The \textit{DFS broker} provides a normalized filesystem interface and translates normalized filesystem requests into native filesystem requests and vice-versa. \textit{Hyperspace} is Hypertable's equivalent to Google’s Chubby service. Hyperspace is a highly available lock manager and provides a filesystem for storing small amounts of metadata. Exclusive or shared locks may be obtained on any created file or directory. High availability is achieved by running in a distributed configuration with replicas running on different physical machines. Consistency is achieved through a distributed consensus protocol. Google refers to Chubby as, "the root of all distributed data structures" which is a good way to think of Hyperspace \cite{hypertable2}

\textbf{Cassandra} Apache Cassandra is a massively scalable open source non-relational database that offers continuous availability, linear scale performance, operational simplicity and easy data distribution across multiple data centers and cloud availability zones. Cassandra was originally developed at Facebook, was open sourced in 2008, and became a top-level Apache project in 2010. Cassandra provides a number of key features and benefits, namely \cite{cass}: \\
- Massively scalable architecture:  a masterless design where all nodes are the same, which provides operational simplicity and easy scale-out.\\
- Active everywhere design: all nodes may be written to and read from.\\
- Linear scale performance: the ability to add nodes without going down produces predictable increases in performance.\\
- Continuous availability: offers redundancy of both data and node function, which eliminate single points of failure and provide constant uptime.\\
- Transparent fault detection and recovery: nodes that fail can easily be restored or replaced.\\
- Flexible and dynamic data model: supports modern data types with fast writes and reads.\\
- Strong data protection: a commit log design ensures no data loss and built in security with backup/restore keeps data protected and safe.\\
- Tunable data consistency: support for strong or eventual data consistency across a widely distributed cluster.\\
- Multi-data center replication: cross data center (in multiple geographies) and multi-cloud availability zone support for writes/reads.\\
- Data compression: data compressed up to 80\% without performance overhead.\\
- CQL (Cassandra Query Language): an SQL-like language that makes moving from a relational database very easy.\\
Cassandra’s architecture is responsible for its ability to scale, perform, and offer continuous uptime. Rather than using a legacy master-slave or a manual and difficult-to-maintain sharded design, Cassandra has a masterless 'ring' architecture that is elegant, easy to set up, and easy to maintain. Cassandra's built-for-scale architecture means that it is capable of handling large amounts of data and thousands of concurrent users or operations per second— even across multiple data centers - as easily as it can manage much smaller amounts of data and user traffic. To add more capacity, new nodes can simply be added to an existing cluster without having to take it down first. Cassandra's architecture also means that, unlike other master-slave or sharded systems, it has no single point of failure and therefore is capable of offering true continuous availability and uptime \cite{cass}.

While Cassandra is a general purpose non-relational database that can be used for a variety of different applications, there are a number of use cases where the database excels over most any other option. These include \cite{cass}:\\
- Internet of things applications: Cassandra is perfect for consuming lots of fast incoming data from devices, sensors and similar mechanisms that exist in many different locations.\\
- Product catalogs and retail apps: Cassandra is the database of choice for many retailers that need durable shopping cart protection, fast product catalog input and lookups, and similar retail app support.\\
- User activity tracking and monitoring: many media and entertainment companies use Cassandra to track and monitor the activity of their users’ interactions with their movies, music, website and online applications.\\
- Messaging: Cassandra serves as the database backbone for numerous mobile phone and messaging providers' applications.\\
- Social media analytics and recommendation engines: many online companies, websites, and social media providers use Cassandra to ingest, analyze, and provide analysis and recommendations to their customers.\\
- Other time-series-based applications: because of Cassandra’s fast write capabilities, wide-row design, and ability to read only the columns needed to satisfy queries, it is well suited time series based applications.\\

It is worth mentioning that Cassandra is a subject of debate when it comes to classification. It can also be classified as a Key-Value Databases, depending on the criteria of the classification (e.g. storage, data model).

\subsubsection{Document-oriented Databases}

Document databases were designed to handle the storage and the management of large scale documents. This type of database assigns a key value to each document. Documents may contain multiple key-value pairs, or key-array pairs, or even nested documents. Documents are encoded in a standard data exchange format such as XML, JSON (Javascript Option Notation) or BSON (Binary JSON). Document databases are recognized as a powerful, flexible and agile tool to store Big Data. Document databases are different from key-value stores. In fact, while key-value stores enable to search for data only by key value, document databases allow users to search for data based on the content of documents. They can query either by keys, values or examples. In fact, the encoded documents contain metadata objects, so it is possible to query data by example. This gives document databases a great flexibility required by some use cases. To launch queries, users can either rely on a programming API or a query language. On the contrary of the simple key-value stores, the value column in document databases contains semi-structured data and specifically the attribute name/value pairs. Furthermore, document databases support a flexible unrestrictive schema. Indeed, they allow storing hundreds of attributes in a single column of a document scheme. So rows can receive various amount and types of attributes \cite{kv2}. Storing data in interpretable JSON documents have the additional advantage of supporting data types, which makes document stores very developer-friendly. Document stores offer multi-attribute lookups on records which may have complete different kinds of key-value pairs. Therefore, these systems are very convient in data integration and schema migration tasks. Most popular use cases are real time analytics, logging and the storage layer of small and flexible websites like blogs \cite{kv1}.

The most prominent Document stores are CouchDB and MongoDB.

\textbf{CouchDB} is a database that completely embraces the web. The data is stored in JSON documents, and accessed with the web browser. It works well with modern web and mobile apps. Data can be distributed efficiently uding CouchDB's incremental replication. CouchDB supports master-master setups with automatic conflict detection. It comes with a suite of features, such as on-the-fly document transformation and real time change notifications. It is highly available and partition-tolerant, and is also eventually consistent. A CouchDB server hosts named databases, which store documents. Each document is uniquely named in the database. Documents are the primary unit of data in CouchDB and consist of any number of fields and attachments. Documents also include metadata that’s maintained by the database system.  Document fields are uniquely named and contain values of varying data (e.g. text, number, boolean, lists, etc.), and there is no set limit to text size or element count. The CouchDB file layout and commitment system features all Atomic Consistent Isolated Durable (ACID) properties. On-disk, CouchDB never overwrites committed data or associated structures, ensuring the database file is always in a consistent state. This is a “crash-only” design where the CouchDB server does not go through a shutdown process, it’s simply terminated. CouchDB is a peer-based distributed database system.  It allows users and servers to access and update the same shared data while disconnected. Those changes can then be replicated bi-directionally later. The  CouchDB  document  storage,  view  and  security  models  are  designed  to  work  together  to  make  true  bidirectional replication efficient and reliable.  Both documents and designs can replicate, allowing full database applications (including application design, logic and data) to be replicated to laptops for offline use, or replicated to servers in remote offices where slow or unreliable connections make sharing data difficult. The replication process is incremental. At the database level, replication only examines documents updated since the last replication. Then for each updated document, only fields and blobs that have changed are replicated across  the network.  If replication fails at any step, due to network problems or crash for example, the next replication restarts at the same document where it left off \cite{couchdb}.

\textbf{MongoDB} is a schemaless document-oriented database. The name Mongo DB comes from the name “humongous”. The database is proposed to be scalable and is written in C++. The main reason for moving away from a relational model is to make scaling easier. MongoDB supports BSON(Binary JSON) data structures to store complex data types and also supports a complex query language. It gives the high-speed admittance to store the mass data. It should store and distribute large binary files like images and videos and instead of stored procedures, developers can store and use JavaScript functions and values on the server side. Then it supports an easy-to-use protocol for storing large files and it gives a fast serial performance for single clients. MongoDB uses memory mapped files system for faster performance. MongoDB holds a set of collections. A collection does not have a pre-defined schema, and stores data as documents. BSON (binary JSON) are used to store documents. A document is a set of fields and can be thought of as a row in a collection. It can hold complex structures such as lists and documents. Each document has an ID field, that is used as a primary key and each collection can hold any kind of document, but queries and indexes can only be made against one collection. MongoDB has supported the indexing over embedded objects and arrays thus have a special feature for arrays called “multi-keys”. That feature allows using the array as an index that can be used to search documents by their linked tags. A MongoDB cluster is built using three main components: Shard nodes, Configuration servers and Routing services or mongos. A Mongo cluster should contain one or more shards, and each shard node is dependable for storing the real data into the database. Each shard consists of either one node or a replicated node, which only holds data for that specific shard. Read/write queries are a retreat to the appropriate shards. A replicated node contains one or more servers. One of the servers acts as the primary server while the other ones are secondary servers. If the primary server fails, one of the secondary ones automatically takes over as primary. All reliable read/write operations go to the primary server and all eventually consistent reads are distributed among all the secondary servers \cite{mongodb}.

\subsubsection{Graph-oriented Databases}

Graph databases are suitable to store not only information about objects but also all relationships existing among them. They rely on a schema-free graph model in order to easily model and represent connected data. Graph models use vertices (e.g., objects or items represented by nodes) and edges to represent connections between data. To illustrate this, a graph can refer to a professional network like the one in Viadeo. In this case, the vertices represent professionals while the directed edges represent links and relationships between those professionals. Each vertex is also initialized with a value. It is worth mentioning that even if graph databases save relationships, they are nothing like relational databases. They are useful to store, access and analyze the strength and the nature of relationships between two or more items (e.g. how close is the relationship between two people? How far away is a taxi driver from another one or from a touristic site). Answering such questions make it possible to formulate valuable recommendations in many industries \cite{Oussou}. 

Graph databases offer for many use cases enhanced performance (they ensure lower latency in comparison with batch processing of aggregates), a flexible data model (easy way to express relationships and to enrich the graph as data and business requirements get more precise) and agility (ability of applications to evolve in a controlled manner aligned with agile and test-driven software development practices). Unlike most classes of NoSQL data stores, graph databases are not the best solution for updating sets of data or for very large volumes of data \cite{robinson}. 

Graph databases are used in social networks, information networks, technological networks (e.g. Internet, airline routes, and telephone networks), biological networks (e.g. area of genomics), and the semantic web. Graph databases are a powerful tool for graph-like queries, for example computing the shortest path between two nodes in the graph. Graph database models are applied in areas where information about data interconnectivity or topology is more important as, or as important as the data itself. Because they are schema-free, graph databases have many advantages amongst which we can cite their natural modeling of data, special graph storage structure, efficient graph algorithms and support for query languages and operators to query the graph structure \cite{angeles}.

A graph contains Nodes and Relationships as mentioned above, the simplest possible graph is a single node, a record that has named values referred to as Properties. A node could start with a single property and grow to a few million. At some point, it makes sense to distribute the data into multiple nodes, organized with explicit Relationships. These relationships organize Nodes int arbitrary structures, allowing a Graph to resemble a List, a Tree, a Map or a compound Entity - any of which can be combined into yet more complex, richly inter-connected structures. Formally, a graph is just a collection of vertices and edges or, in less intimidating language, a set of nodes and the relationships that connect them. Graphs represent entities as nodes and the ways in which those entities relate to the world as relationships. The simplest possible graph is a single node, a record that has named values referred to as Properties. A node could start with a single property and grow to a few million. At some point, it makes sense to distribute the data into multiple nodes, organized with explicit Relationships. Relationships organize Nodes into arbitrary structures, allowing a Graph to resemble a List, a Tree, a Map, or a compound Entity – any of which can be combined into yet more complex, richly inter-connected structures. This general-purpose, expressive structure allows the modeling of all kinds of scenarios, from the construction of a space rocket, to a system of roads, and from the supply-chain or provenance of foodstuff, to medical history for populations, and beyond \cite{robinson}, \cite{shimpi}.

Graphs are extremely useful in understanding a wide diversity of datasets in fields such as science, government, and business. The real world—unlike the forms-based model behind the relational database—is rich and interrelated: uniform and rule-bound in parts, exceptional and irregular in others. Once we understand graphs, we begin to see them in all sorts of places. Gartner, for example, identifies five graphs in the world of business—social, intent, consumption, interest, and mobile—and says that the ability to leverage these graphs provides a “sustainable competitive advantage.” For example, Twitter’s data is easily represented as a graph. For example, Twitter’s data is easily represented as a graph. The Figure below shows a small network of Twitter users. Each node is labeled User, indicating its role in the network. These nodes are then connected with relationships, which help further establish the semantic context: namely, that Billy follows Harry, and that Harry, in turn, follows Billy. Ruth and Harry likewise follow each other, but sadly, although Ruth follows Billy, Billy has not (yet) reciprocated. Of course, Twitter’s real graph is hundreds of millions of times larger than the example, but it works on precisely the same principles \cite{robinson}.\\

\textbf{VertexDB} is a high performance graph database server that supports automatic garbage collection. It uses the HTTP protocol for requests and Java Script Object Notation( JSON) for its response data format and the API are inspired by the FUSE file system API plus a few extra methods for queries and queues. Vertexdb is used as a graph store \cite{shimpi}.

Vertexdb’s data model is constituted as follows \cite{vertexdb}: Nodes are collections of lexically ordered key/value pairs; A value can contain either data or a pointer to another node; Keys for data begin with an underscore.

Vertexdb uses an asynchronous I/O system, small requests – which it handles serially, periodic backups and garbage collection \cite{vertexdb}. 

\textbf{Infinitegraph} is a distributed object database with C++, java, C\# and python bindings. Infinitegraph’s architecture is constituted of four layers, namely: User application, API, Management and configuration sections and Objectivity and distributed databases. Infinitegraph is used to extend business, social and government intelligence with
graph analysis, and is built on a highly scalable, distributed database architecture where both data and processing are distributed across the network. It can handle large transactions efficiently. In organizations infinitegraph is used to identify and understand complex relationships between distributed data \cite{shimpi}.\\

\textbf{Hypergraphdb} \cite{iordanov} is a general purpose, extensible, portable, distributed, embeddable, open-source data storage mechanism. Although a hypergraph database is mostly designed
for knowledge representation, Artificial Intelligence (AI) and semantic web projects, it can also be used as an embedded object-oriented database for Java projects of all sizes. The architecture of Hypergraphdb is composed of various layers, namely: Applications, Querying, Model layer, Primitive storage layer and Key-value store. Hypergraphdb is useful for areas like knowledge representation, artificial intelligence and bio-informatics \cite{shimpi}.

\textbf{Sones GraphDB} was a graph database developed by the german company Sones GmbH. It was available from 2010 to 2012, the last version was released in May 2011. The company went bankrupt on January 1st ,2012. Sones GraphDB was unique with a design based on weighted graphs, and had a modular structure consisting of 4 application layers. The storage engines acted as the interface to different storage media. The GraphFS serialized and deserialized database objects (nodes and edges) and operated the available storage engines. The actual graph-oriented database logic as well as all functionalities specific to the database was implemented in the GraphDB. The GraphDS provided the interface for using the database. The interfaces between the application layers were generic, which made it possible to update components separately. This graph database provided an inherent support for high-level data abstraction concepts for graphs. It defined its own query language and an underlying distributed file system \cite{shimpi}.

\textbf{Infogrid} is an open-source web graph database, developed in Java and whose functions are oriented to web applications. It can be used as a standalone graph database or in addition to the other InfoGrid projects. InfoGrid Graph Database (Grid) augments the graph database with a replication protocol, so that many distributed Graph Databases can collaborate in managing very large graphs. InfoGrid stores provide an abstract common interface to storage technologies such as SQL databases and distributed NoSQL hash tables. InfoGrid User Interface REST-fully maps the content of a Grap hDatabase to browser-accessible URLs. Viewlets allow developers to define how individual objects and sub-graphs are rendered. The project also implements a library of Viewlets, and the MeshWorld and NetMeshWorld example applications. InfoGrid Light-Weight Identity project implements user-centric identity technologies such as LID and OpenID. InfoGrid Model Library Project defines a library of reusable object models that can be used as schemas for InfoGrid applications. InfoGrid Probe Project implements the Probe Framework, which enables application developers to treat any data source on the Internet as a graph of objects. InfoGrid Utilities project collects common object frameworks and utility code used throughout InfoGrid \cite{shimpi}.

\textbf{Neo4j} is a high-performance, NoSQL graph database with all the features of a mature and robust database. Neo4j is debatably the most popular graph database. It is particularly developed for Java applications, but it also supports Python. It is an open source project available in a General Public License (GPL) v3 Community edition, with Advanced and Enterprise editions available under both the Advanced GPL (AGPL) v3 as well as a commercial license. The graph model in Neo4j consists of three characteristics, namely: a Property (a key-value pair that can be added to both nodes and edges), Only edges can be associated with a type (e.g. “KNOWS”) and Edges can be specified as directed or undirected. Neo4j uses the following index mechanism: a super referenceNode is connected to all the nodes by a special edge type “REFERENCE”. This actually allows to create multiple indexes to distinguish them by different edge types. Neo4j implements an object-oriented API, a native disk-based storage manager for graphs, and a framework for graph traversals \cite{shimpi}.

\subsubsection{Disambiguation of NoSQL Databases}

As shown in the previous sections, NoSQL databases are classified, by consensus, into four major types. There are many misconceptions as well as disagreements around this classification and the databases included in it.

A major disagreement revolves around including Graph databases into the NoSQL world. Many designers consider Graph databases to be inherently different from the three other types of NoSQL. The Graph data model explicitly describes the relationships and dependencies between data nodes, while other types of databases (including relational databases) do so implicitly. The rich theoretical background of graph databases stemming from graph theory makes it that much different from other industry-born databases.

Due to their distinct characteristics, many designers consider graph databases to be a separate subject in their own right, and thus should not be 'lumped in' with a group of databases it is not a part of \cite{bloor}. The example of Neo4j is often used to state that even though it supports ACID-compliant transactions and XA-compliant two phase commit, it is also geared towards handling significant query processing and offers a very flexible design \cite{bloor2}. These characteristics are sometimes falsely used to justify that some Neo4j (amongst other NoSQL Databases) should fall under the NewSQL umbrella. We find this inference wrong, given that NewSQL is a relational class of databases, and that being ACID-compliant is not enough of a criterion to justify not being a NoSQL database. Many NoSQL databases now offer ACID transactions. For instance, MarkLogic demonstrates in their white paper how they manage ACID support \cite{marklogic}. As the Product VP of yagabyteDB highlights in \cite{signs}, ACID compliance does not require SQL, and new research from Google (Google Spanner, Percolater) and from Yale (Calvin) are "serving as inspiration for a new generation of distributed ACID compliant databases.  

Another problematic ambiguity concerning NoSQL databases is the confusion between Column-Family databases and Columnar databases. The former being NoSQL databases and the latter relational. Abadi makes, in \cite{abadicolumnar}, the remark  that Bigtable, HBase, Hypertable, and Cassandra are being called column-stores with increasing frequency due to their ability to store and access column families separately. This connotation has created a confusion with relational column stores such as  Sybase IQ, C-Store, Vertica, VectorWise, MonetDB, ParAccel, and Infobright which are also able to access columns separately.

Abadi highlights the differences between these two groups stating that the data model of Column Family databases is basically a sparse distributed, persistent multi-dimensional sorted map, whereas Columnar databases follow a traditional relational model. He also adds independence of columns, SQL support, storage layer and optimized workloads as major distinctions between the two groups. In the same blogpost, Abadi posts an adendum citing Stonebraker's statement that column-family databases are "row stores" which store a column family in a row-by-row fashion, use materialized views for column families and store materialized views row-by-row, whereas relational columnar databases have a sophisticated column-oriented optimizer". The subtle differences arise even when making the distinction between relational row-oriented and relational columnar databases. A common ambiguity is the perception that a column store is merely a row-store with an index on every column. The difference lies in the mapping of the data. In a row-oriented indexed system, the primary key is the 'row id' that is mapped from indexed data. In the column-oriented system, the primary key is the data, which is mapped from row ids \cite{debunk}.
\subsection{CAP-based Classification of NoSQL Databases}
To reiterate what has been aforementioned, the CAP theorem outlines a sort of road map into how individual systems handle network partitions, whether they will prioritize consistency or availability, and to which degree. The CAP theorem is often used to classify various NoSQL systems based on the out-dated notion of two-out-of-three. In what follows, we discuss -- based on technical benchmarks and research, how a few NoSQL databases can be accurately and correctly classified using CAP.

\textbf{MongoDB} is a single-master system. Each replica set can have only one primary node that receives all the write operations. All other nodes in the same replica set are secondary nodes that replicate the primary node's operation log and apply it to their own data set. By default, clients also read from the primary node, but they can also specify a read preference that allows them to read from secondary nodes. When the primary node becomes unavailable, the secondary node with the most recent operation log will be elected as the new primary node. Once all the other secondary nodes catch up with the new master, the cluster becomes available again. As clients cannot make any write requests during this interval, the data remains consistent across the entire network \cite{ibm}.

MongoDB manages its high horizontal scalability using automatic sharding. \cite{consist} explains how MongoDB follows ACID properties by: supporting single operation inserts and updates; supporting the possibility of strong consistency; managing isolation of documents while they are being updated; using 'write concern' as a user defined policy to be fulfilled before a commit.

MongoDB allows configuring a replica set. A replica set has primary replica set members and secondary replica set members. There are two configurations based on the desired consistency level: Strong consistency where applications write and read from the primary replica set member, and Eventual consistency where applications can read from secondary replica set members if they do not prioritize reading the latest data \cite{consist}.

\textbf{Neo4j} provides a fault tolerant platform, reading scale up and a Causal Consistency model. A cluster is composed of two types of nodes, core servers and read replicas. In a single node architecture, Neo4j is strongly consistent. In Neo4j Enterprise Edition, the cluster ensures causal consistency which guarantees that reading data previously written from the same client will be consistent. However, eventual consistency is obtained when reading data that was changed by other clients, because there is a millisecond time window which the latest data has not been propagated yet \cite{consist}.

\textbf{Cassandra} distributes data across a set of nodes, designated as a cluster, thus offering scalability. Cassandra partitions data across the cluster by hashing the row key. To achieve a highly available system, Cassandra uses replication. Cassandra was initially designed to be eventually consistent, high-available and low-latency. However, its (read and write) consistency can be tuned to match the client’s requirements. On the default configuration, Cassandra updates all the replica nodes that have been queried in some reading request to reflect the latest value. This routine is called Read Repair and, because a single read triggers it, it puts little stress on the cluster \cite{consist}. 

\textbf{Redis} implements a master-slave model without proxies which means that the application is redirected to the node that has the requested data. Redis nodes do not intermediate responses. On the asynchronous replication configuration (default), if the master node dies before replicating and after acknowledging the client, the data is permanently lost. Therefore, the Redis Cluster is not
able to guarantee write persistence at all times \cite{consist}. There are various scenarios which \cite{consist} highlights to demonstrate how each of them affects (eventual) consistency in Redis.

In their study on consistency models in NoSQL databases, \cite{consist} concluded that configuring any selected database
to favor strong consistency will result in less availability when subject to network partition events (as stated by the CAP theorem). They also inferred based on the tests they conducted that MongoDB is the preferable option to ensure high consistency in a fault-tolerant environment, while Cassandra is the better choice to provide high availability. 

A new branch of databases has been emerging for a while, following in the steps of polyglot persistence: Multimodel databases. It is worth noting that the main difference is that, in a polyglot persistent environment, multiple data models are supported using multiple data stores, while a multimodel database uses a single data store to support multiple data models. In light of these novel alternatives, including the rise of NewSQL, it is crucial to explore how CAP affects different systems, and how designers can maximize its properties for their respective applications.

\section{Discussion}

Although many critics claim that the CAP theorem is an impossibility result which appeals to theorists but fails to capture the complexity of designing distributed systems, it still represents a cornerstone of distributed database theory. We believe that although it has been misunderstood and as Brewer himself states "abused", it is an important and proven result which remains verified and is seriously taken into consideration and needed in database design.

The fact that it fails to include all possible errors and singles out network partitions is, in our opinion, not a weakness. Instead, we consider it to be a direct result of a mathematical formalization which inevitably limits its area of application. It does not make it any less true or any less efficient. The particularity of network partitions is that they present a case in which the nodes are still accessible, which means that our system suffers from a communication breakdown splitting it into partitions which can either be available or consistent. Other types of network failure or database errors would have consequences on the consistency and availability of the database, but would not create an impossibility result the way the CAP theorem does. In \cite{brewer12}, Brewer states that the 'old' two-out-of-three expression of CAP served its purpose of opening minds of to a wider range of systems and tradeoffs. He also highlights the importance of exploring the nuances required to push the traditional way of dealing with partitions, and explains that the goal of the modern CAP should be to maximize combinations of consistency and availability that make sense for the specific application.

Modern systems should take advantage of the various theoretical deductions and technical findings. In general, real systems (e.g. The Cloud) use both ACID and Base. Analogically, the aim should be to use the CAP theorem correctly at the design phase, all while taking into consideration the wide range of flexibility that novel systems provide. These novel systems (NoSQL, NewSQL or otherwise) do not always have to 'reinvent the wheel' and completely start over. Hybridization of 'old' and new systems and concepts can lead to revolutionizing both the industrial and research worlds. Thoretical foundations are as important as technological advances in the NoSQL world, especially now that the NoSQL community not only yearns for disambiguation but also seeks standardization and unification of design methodologies as specified in \cite{asaad2018}.

\section{Conclusion}
NoSQL Databases are highly scalable, heterogeneous, numerous, Big-Data-oriented. They provide a flexible and schema-less design alternative to the traditional one-size-fits-all paradigm, and are clustered under four big families: key-value, column-family, document and graph databases. The theoretical foundation of NoSQL -- or rather lack thereof, left a gap in the literature and created major debates and difficulty of achieving consensus in even the most 'basic' areas of NoSQL. The CAP theorem, the inclusion of Graph databases, the differences between columnar databases and NoSQL column-family databases, and the classification of NoSQL databases are but some of the many ambiguous and misinterpreted areas of NoSQL. Disambiguating these problematic misconceptions is the first step in standardizing and unifying the NoSQL world in research and industry alike.

\bibliographystyle{unsrt}  


\end{document}